\documentstyle[11pt,newpasp,twoside,epsf]{article}
\def\edcomment#1{\iffalse\marginpar{\raggedright\sl#1\/}\else\relax\fi}
\reversemarginpar

\begin{document}
\title{Beyond the Cosmic Veil: 
Star and Planet Formation Research after SIRTF and NGST}
 \author{Michael R. Meyer}
\affil{Steward Observatory, The University of Arizona, \\
933 N. Cherry Avenue, Tucson, AZ 85721}

\begin{abstract}
How do planets form from circumstellar disks of gas and dust? 
What physical processes are responsible for determining the final masses 
of forming stars and ultimately the initial mass function (IMF)? 
The Hubble Space Telescope has made major contributions in helping to 
address these fundamental questions.  In the next decade, the Space
Infrared Telescope Facility (SIRTF) and the Next Generation Space 
Telescope (NGST) will build on this heritage in the near-- to 
far--infrared.  However several crucial questions will remain. 
We review recent progress made in star and planet formation with 
HST, summarize key science objectives for SIRTF and NGST, and suggest 
problems that would be uniquely suited to a large aperture UV/optical 
space--based telescope.  We focus on studies that take 
advantage of high spatial resolution and the unique 
wavelength range not accessible from the ground such as: 
1) circumstellar
disk structure and composition (resolved images
of dust and UV spectroscopy of gas); and 2) extreme 
populations of young stars in the local group (UV imaging 
and spectroscopy of massive star--forming regions). 
An 8m UV/O space telescope
operating from 1100--6000 $\AA$ over fields of view 4--10'
and with spectroscopic capabilities from R = 3,000--10,000 
down to $<$ 1000 $\AA$ 
would be a powerful tool for star and planet formation research. 
\end{abstract}

\section{Introduction}

Star and planet formation will remain key themes in the NASA Origins
Program through the next decade and beyond.  During this time, the 
community will enjoy access to: 1) continued HST operation with 
STIS, NICMOS, ACS, COS, and WFC--3 with wavelength coverage from 
0.1150--2.5 $\mu$m; 2) SIRTF with launch in early 2003 (lifetime 
$\sim$ 5 years) covering 
the wavelength range from 3--160 $\mu$m; and 3) NGST scheduled
for launch in 2010 with projected 5 year mission covering the
wavelength range 0.6--28 $\mu$m.  Because important processes 
in star and planet formation occur over a wide range of temperatures, 
multi--wavelength observations are required in order to make significant progress
(see overview by Hartmann this volume). 

Both SIRTF and NGST will execute programs aimed at 
understanding 
the emergence of planetary systems from circumstellar
disks around young stars, and 
the origins of stellar masses. 
Approved SIRTF guaranteed time 
programs will focus on circumstellar disk evolution
and the nature and frequency of brown dwarf objects. 
In addition, three of the six adopted Legacy Science Programs 
(GLIMPSE, From Molecular Clores to Planet--Forming Disks, 
Formation and Evolution of Planetary Systems) 
will directly impact star and planet formation research 
(http://sirtf.caltech.edu/SSC/legacy/).  Two of 
the five themes in the Design Reference Mission for the NGST 
(The Birth and Formation of Stars and The Origins and Evolution
of Planetary Systems) deal with similar topics 
(http://ngst.gsfc.nasa.gov/science/drm.html). 

Observations in the UV/optical range are also important to 
obtain a complete picture of star and planet formation. 
Key observations include: 1) sampling the
Wien peak of massive star energy distributions; 2) measuring important gas 
diagnostics for infall/outflow; and 3) studying 
resolved images of dust disks.  As in the infrared spectral regime, space
offers unique capabilities for UV/optical astronomy such as 
the accessible range from 912--3000 $\AA$, and diffraction--limited
imaging over a large field of view at wavelengths $<$ 1 $\mu$m. 

Here, we concentrate on two areas where 
a large aperture UV/Optical space telescope could
make substantial contributions.  We begin by 
identifying key issues, review the legacy of HST 
in each area, summarize the promise of SIRTF for 
studies of circumstellar disks and 
NGST for investigations concerning the origin of 
the IMF, and suggest parameters required
in order for a UV/optical space telescope to realize
its potential for star and planet formation research. 

\section{Evolution of Circumstellar Disks}

Why should we care about the evolution
of circumstellar disks around young stars?  
Because their study holds the promise of 
connecting observations of disks 
observed in the Milky Way as a function of
age with the origin and evolution of our 
own solar system.  Disks are the 
mostly likely sites of planet formation
giving rise to extra--solar planets
such as those detected around sun--like stars in the 
solar neighborhood.  Ultimately we 
wish to know whether planetary 
systems are common or rare in the Universe.  
In addition, disk accretion is an 
important part of the star formation
process.  Does it help provide ``feedback''
into the surrounding interstellar medium 
and thus a regulating mechanism for 
the determination of stellar masses?
What effects does it have 
on pre--main sequence evolution? 

Our current understanding of the evolution
of circumstellar disks surrounding young stars
is limited by available observations.  High resolution 
spectroscopic observations have suggested a connection
between the accretion of material through a disk 
and mass loss in powerful outflows (e.g. Edwards et al. 1994). 
Through ground--based near--IR studies of young clusters, it appears
that inner disk accretion terminates on timescales
$<$ 10 Myr for most stars (Haisch et al. 2001; Hillenbrand et al. 2002). 
Photometric monitoring and 
spectroscopic observations have suggested a link 
between stellar rotation and the presence or absence of a disk 
(e.g. Bouvier et al. 1993).  
Studies of outer disks 
(1--10 AU) require mid-- to far--infrared
observations.  A number of groups have suggested that
substantial evolution occurs in dust disks at 
these radii from 10--100 Myr (see Meyer \& Beckwith 2000 for a 
recent review).  Millimeter
wave observations probe evolution of the coolest
dust found at the largest radii (e.g. Zuckerman \& Becklin 1993). 

\subsection{Legacy of Hubble}

The Hubble Space Telescope (HST) has provided a rich legacy of 
imaging and spectroscopic observations that have revolutionized
the study of circumstellar disks around young stars.  The 
first images of the dark disks seen in 
silhouette against the bright background of the Orion 
Nebula provided direct confirmation of the disk hypothesis
that even the most ardent skeptics find difficult to 
refute (e.g. McCaughrean \& O'Dell 1996; Figure 1). 
Subsequent studies have offered numerous examples of 
objects seen edge--on where a favorable viewing 
geometry provides a wealth of detailed knowledge 
about individual objects (e.g. Burrows et al. 1996). 
Monitoring programs
over several years have also enabled proper 
motion studies of accretion--powered jets that provide 
direct tests of theoretical models (Hartigan et al. 2001).  Coronographic 
observations with STIS 
and NICMOS (Schneider \& Silverstone this volume) utilize
high contrast imaging as a tool to probe dust 
structure and composition.  Wide--field imaging
studies have provided insight concerning 
large--scale mass--loss and interactions with 
the surrounding interstellar medium (Bally et al. 2002).   
Finally, UV spectroscopy
is proving to be a powerful diagnostic tool to 
study accretion processes (Ardila et al. 2002)
and the atomic and molecular gas 
content of disks through absorption--line 
observations along favorable lines of sight 
(e.g. Vidal--Madjar et al. 1994). 
Long--slit diffraction--limited imaging spectroscopy has also shed light on 
the complex relationships between star/disk systems and the surrounding 
interstellar medium (Grady et al. this volume). 

\begin{figure}
\plottwo{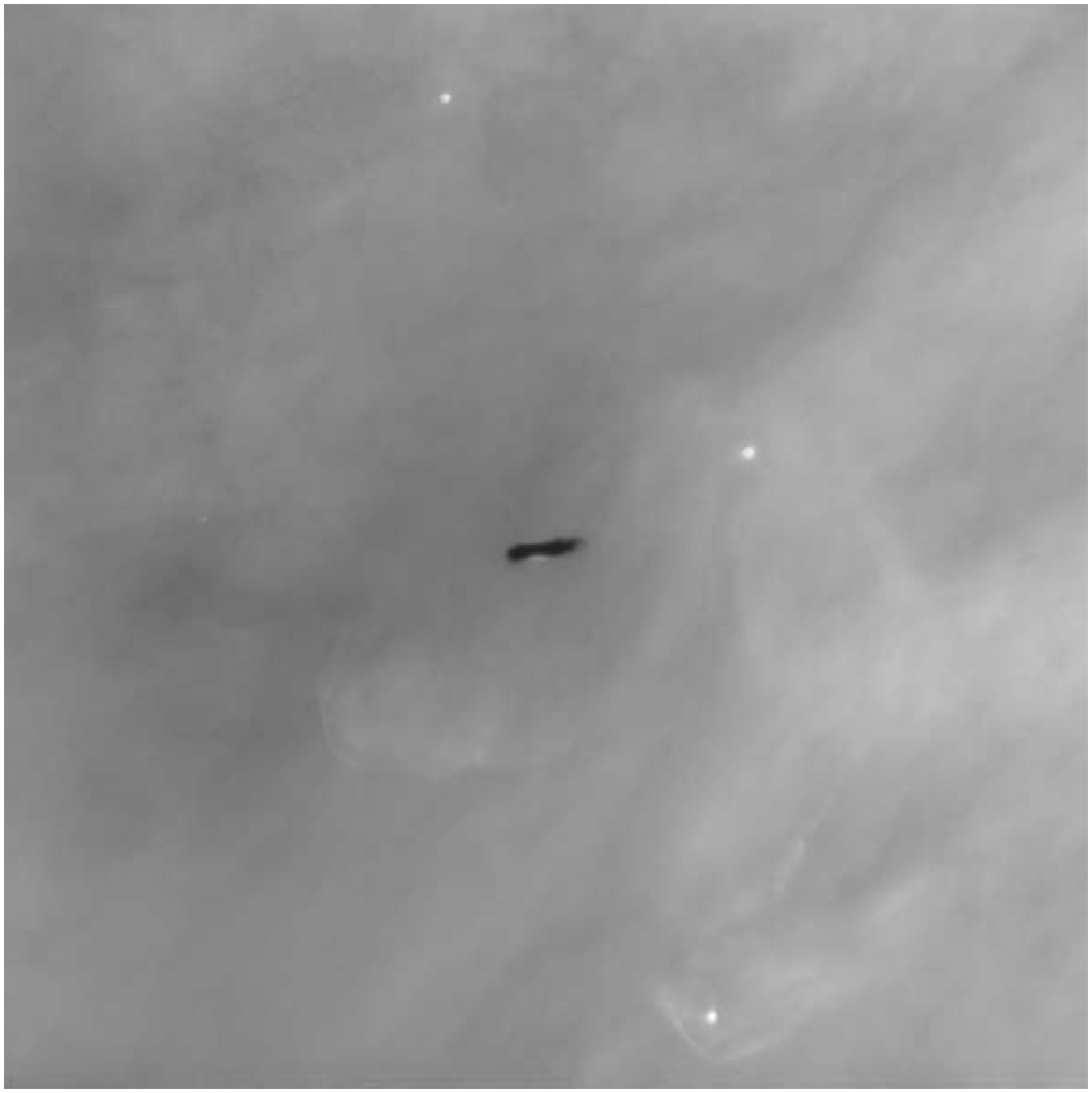}{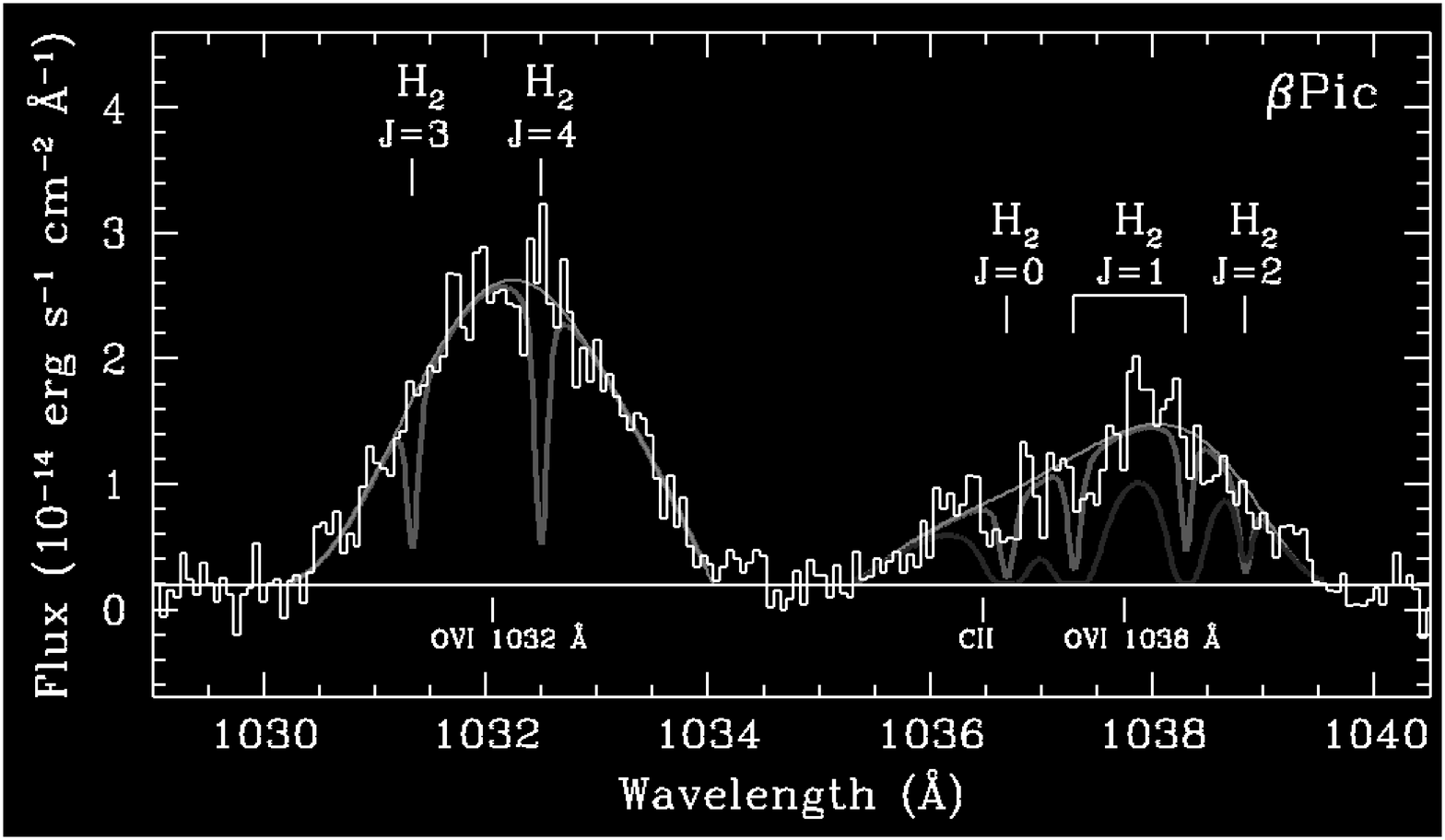} 
\caption{Left:  WFPC2 image of an edge--on disk seen in 
silhouette against the emission--line background of the Orion 
Nebula (Credit: NASA, J. Bally, H. Throop, and C.R. O'Dell).  
Right: FUSE observation of $\beta$ Pictoris 
from Lecavelier des Etangs et al. (2001).  The dotted line is a 
model for cold molecular hydrogen gas absorption seen against the 
continuum provided by OVI emission lines.}
\end{figure} 

\subsection{The Promise of SIRTF} 

The Space Infrared Telescope, the last of NASA's 
``Great Observatories'' and the first 
new mission of the NASA {\it Origins} program, 
will be launched in early 2003 and provide 
unprecedented mid-- and far--infrared sensitivity
for studies of circumstellar disks.  Through 
a suite of guaranteed time programs and the 
Legacy Science Program, 
SIRTF will make major contributions to 
our understanding of the structure and 
evolution of circumstellar disks around 
young stars.  One program, the 
Formation and Evolution of Planetary Systems
(Meyer et al. 2002), 
is aimed directly at studying 
dust disks around solar--type stars and placing 
our solar system in context
\footnote{For additional information see http://feps.as.arizona.edu.}. 
Through construction of 
spectral energy distributions from 3--160 $\mu$m 
for 350 stars of spectral type F8--K3 we hope to: 
1) characterize 
the transition from primordial dust disks
to debris disks by tracing evolution in the 
amount, distribution, and composition of dust; and 2) examine the diversity 
of planetary systems through their
dynamical effect on dust.  Through spectroscopic
observations of warm (50--200 K) molecular hydrogen  
gas we hope to constrain the timescale
for gas disk dissipation and giant planet formation. 
These programs, which build directly on the 
heritage of IRAS and ISO, will leave a rich 
legacy for follow--up observations with 
NGST as well as a large aperture 
UV/optical space telescope. 

\subsection{Requirements for UV/O Space Telescope} 

What should we require from a UV/optical space 
telescope (UV/O--ST) for studies of planet formation? 
Spatial resolution of order 0.3 AU for the nearest
T Tauri stars (3 mas at 100 pc and 1100 $\AA$) would enable us to:
1) possibly resolve the disk/jet interface providing
a detailed understanding of the accretion process
and insight into the star/disk interaction 
that may mediate stellar angular momentum evolution; 
and 2) detect large gaps in disks created through the 
dynamical interaction of giant planets with dust
debris.  In addition to spatial resolution, 
sensitivity to study gas disks at R = 10,000 is
required in order to: 1) measure accretion rates
$<$ 10$^{-10}$ M$_{\odot}$/year (c.f. Johns--Krull et al. 
2000) 
and 2) extend limits on H$_2$ and CO gas in disks 
to levels comparable to SIRTF (c.f. Figure 1). 
Finally, 4--10' fields of view from 1100--6000 $\AA$
are required to trace the accretion--driven mass 
loss history of young stars 
by observing shock diagnostics over a 
range of velocities.  
Such studies can constrain: 1) the angular
momentum loss of PMS star/disk systems 
through energetic winds; and 2) the 
deposition of kinetic energy into the turbulent
interstellar medium providing a support mechanism 
for collapsing clouds.  
An 8m UV/O--ST meets these requirements. 

\section{The Origins of Stellar Masses}

How did the universal star formation rate evolve over cosmic
time?  How do galaxies, including our own Milky Way, assemble themselves?
How are stars and planets formed?  In order to quantitatively address all of these issues, one
fundamental question must be answered:  what physical processes
determine the shape of the initial mass function of stars and sub--stellar objects?
This question has plagued astronomers since the initial mass function 
was first introduced by Salpeter (1955).  
Its construction requires translation of an observed luminosity function into
a mass function through adoption of an appropriate mass--luminosity (M--L) relation.
Because high mass stars exist on the mass sequence for shorter lifetimes and
because they are rare in comparison to lower mass stars,
even construction of the field star IMF requires the assumption that: 1) the
IMF is the same everywhere; and 2) it has remained the same for all time.
These two assumptions have not been tested over the full range of 
available star--forming environments. 
Recent work by Massey et al. (1995), Scalo (1998), Reid et al. (1999),
Kroupa (2001) and references therein have led to extensions of the field star IMF into new mass regimes
as well as refinements in the characterization of its shape. 
Yet, even with the strong assumptions outlined above, uncertainties
in the galactic birthrate complicate corrections for massive star
evolution and adoption of time--dependent M--L relationships for sub-stellar
objects in well--mixed stellar populations.  Newly formed star clusters
are ideal laboratories to investigate the IMF over the full range of stellar
and sub--stellar masses with a minimum of assumptions.

\subsection{Legacy of Hubble}

Through observations that have complemented ground--based work, 
HST has provided intriguing hints concerning answers to the above 
questions.   Studies of mass functions with WFPC2 and NICMOS in the field 
(Gould et al. 1997), the bulge (Holtzman et al. 1998), 
 spheroid (Gould et al. 1998) and globular clusters 
(Paresce \& De Marchi 2000) 
all point toward a stellar IMF that is similar in a variety of 
star--forming environments. 
Observations of young clusters with NICMOS, 
where brown dwarfs are more easily 
detected due to their high ``pre--main sequence'' luminosities, 
suggest that the universality of the IMF extends into the 
sub--stellar range (Luhman et al. 2000; Najita, Teide, \& Carr 2001). 
Massey \& Hunter (1998) have utilized the FOS to study the IMF 
in the low metallicity 
R 136 cluster in the Large Magellanic Cloud (Figure 2). 
The rapid post--main sequence evolution of massive stars creates
degeneracies in the analysis of color--magnitude diagrams that
require classification spectroscopy in order to accurately derive
the IMF.  Massey and collaborators find that the IMF above 10 M$_{\odot}$ 
does not vary over a range of 2.0 dex in metallicity. 
However, recent near--infrared photometric studies
of massive star formation in the inner galaxy with NICMOS on HST 
have raised the possibility of an unusual power--law IMF above 
5 M$_{\odot}$ in the heavily embedded Arches cluster (Figer et al. 1999). 

\subsection{The Promise of NGST}

Does the Initial Mass Function (IMF) truncate
somewhere below the hydrogen burning limit?  Theoretical speculations abound.
One conjecture is that the star--forming process is essentially {\it self--regulating}
through some negative feedback mechanism.  For example, accretion of material onto the
central protostar drives a powerful bi--polar outflow which could disrupt the
remnant infalling envelope (Adams \& Fatuzzo, 1996).
Balancing the mechanical energy
between infall and outflow for typical molecular cloud cores
suggests a characteristic mass of $\sim$ 0.25 M$_{\odot}$.   An opposing view holds that
the distribution of star--forming units in a collapsing cloud core is determined
by fragmentation processes which in turn depend on the {\it initial conditions} of the
local interstellar medium (e.g. temperature, density, metallicity).  For example,
the mass scale for a gravitational perturbation to become unstable in a
uniform density medium, the Jean's Mass, is roughly 0.7 M$_{\odot}$ when averaged
over the observed range of densities in nearby molecular clouds. Indeed such a preferred
scale is observed as a break in the power--law distributions that characterize the
surface density of companions in the Taurus dark cloud (Larson 1995).
As the fragments collapse, the local gas density increases.  Provided the material
remains at a constant temperature, the Jean's mass will decrease driving smaller and
smaller regions unstable (hierarchical fragmentation).
However, the minimum mass for fragmentation is fixed at the point where the condensing core
becomes opaque to its own radiation, preventing further sub--fragmentation at a constant
temperature (e.g. Spitzer 1978).
Under conditions typical of nearby clouds this limit is
$\sim 10 M_{JUP}$.   Large spectroscopic surveys will be needed
to build robust estimates of the IMF and distinguish between these theories. 
With its combination of high angular resolution, accessible wavelength range, 
 and sensitivity, NGST will be able to obtain NIR spectra for hundreds of 10$^6$ yr 
objects down to 1 M$_{JUP}$ viewed through $A_V < 25^m$ of extinction 
within 500 pc of the Sun in a few hours of integration. 

Nearby regions of massive star formation (R 136, NGC 3603, the Arches cluster) 
have commanded considerable attention as local
analogs to ``super star clusters'' observed in interacting
galaxies (e.g. O'Connell et al. 1994) that are proximate enough for detailed studies of the resolved
stellar populations.  With few exceptions, most studies
show that the IMF of these regions of extreme star formation are consistent with
the field star IMF that characterizes the solar neighborhood down to varying
low mass limits $>$ 1M$_{\odot}$.  Each of these regions contains dozens of
stars $>$ 10 M$_{\odot}$ which according to the field star IMF should be accompanied
by $\sim$ 10$^5$ stars down to the hydrogen burning limit. 
Yet very little is known concerning the
IMF in regions of extreme star formation below 1.0 M$_{\odot}$.  
Theory suggests that lower metallicity regions might contain hotter
gas due to a lack of molecular cooling and therefore exhibit higher Jean's
mass on average (Nakamura \& Umemura 2002). 
The starburst phenomenon is often associated with a top--heavy IMF
as well as an upper mass cutoff suggested by models of their spectral
evolution.  In order to constrain the mass--to--light ratio in
unresolved stellar populations over cosmic time, it is crucial to understand
the shape of the IMF as a function of metallicity and environment
in regions of
extreme star formation.  Our best hope is to study in detail
star--forming events  in the Milky Way
and the local group that approach the activity of starbursts.
NGST will be able to obtain spectra for objects below the 
hydrogen burning limit in clusters aged 1--3 Myr with 
A$_V$ = 0--10 from the Galactic Center out to the distance of the LMC. 

\subsection{Requirements for UV/O Space Telescope}

What capabilities are required for an UV/O--ST in order
for it to make major advances in understanding the 
origin of the initial mass function?  Maintaining a 
wavelength range from 1100--6000 $\AA$ will enable a 
UV/O--ST to study the formation of high mass stars 
in regions of low extinction throughout the local group over 
a wide range of stellar temperatures (spectral types later than 
K0 to earlier than O5).  Spatial resolutions of 
200 AU (3 mas at the distance of the LMC and 1100 $\AA$) will result
in resolution of many wide binary systems and minimize
the nebular background often associated with regions
of massive star formation.  Spectral resolution (and 
sensitivity) at R$>$ 3000 is required for classification
of the hottest stars, a necessary step in order to derive 
an accurate IMF in regions of extreme star formation.  
Fields of view 4--10' are 
required in order to separate clusters from field populations 
and assess local environments which could influence their 
dynamical evolution. 
An 8m UV/O--ST would enable studies of nearby starbursts
with resolution and sensitivity comparable
to HST studies of massive star formation in the LMC (Figure 2). 

\begin{figure}
\plottwo{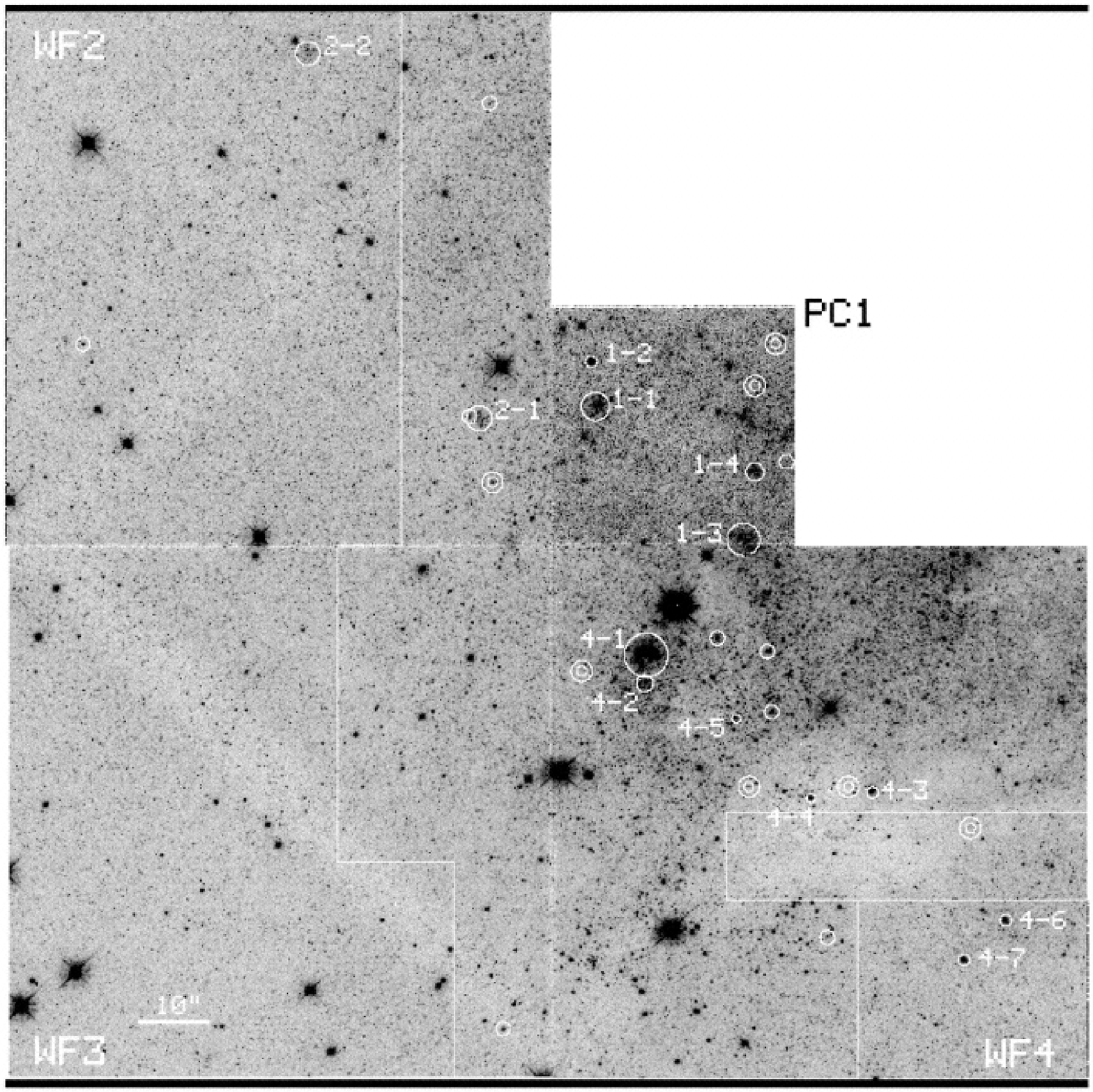}{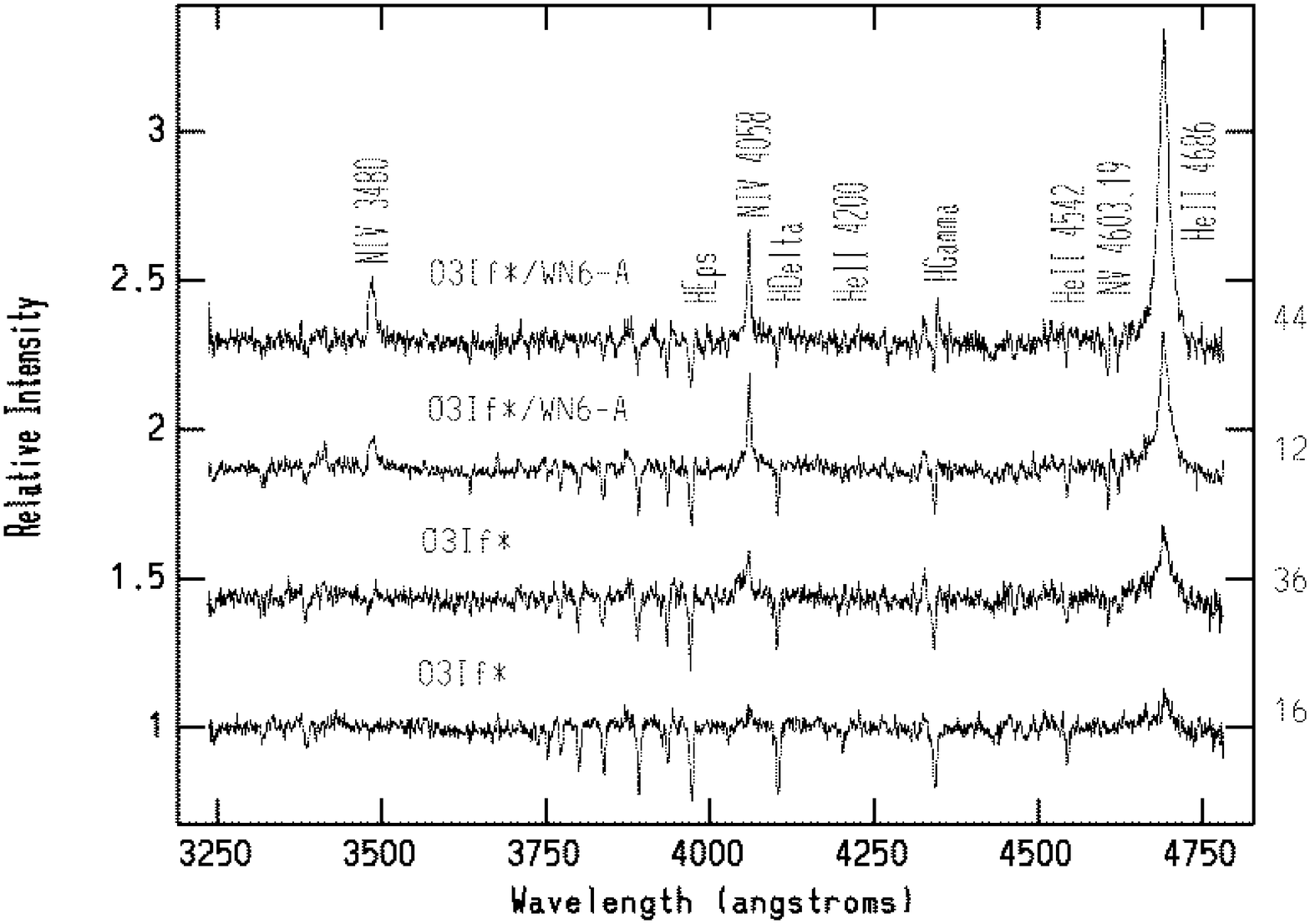} 
\caption{Left:  WFPC2 image of the nearby starburst IC 10 from 
Hunter (2001) with individual star clusters and W--R candidates 
marked.  Right: FOS spectra of massive stars
in the R 136 cluster of the LMC from Massey \& Hunter (1998).  
With an 8m UV/O--ST, studies of the IMF in IC 10 would be 
comparable to extant studies of the LMC with HST.} 
\end{figure} 

\begin{acknowledgements} 
I would like to thank the SOC 
for organizing such a stimulating meeting. 
Special thanks to 
C. Grady, P. Hartigan, P. Massey, and K. Stapelfeldt for discussions
concerning priorities for a UV/O--ST in star and planet
formation research, and M. McCaughrean
for helpful comments on a draft of this manuscript. 
Thanks also to the the Max--Planck--Institut f\"ur Astronomie
in Heidelberg and the Astrophysikalishes Institut Potsdam
for the hospitality they provided during the preparation of this manuscript. 
\end{acknowledgements}

\end{document}